# Decoupling absorption and emission processes in super-resolution localization of emitters in a plasmonic hotspot

David L. Mack[1], Emiliano Cortés[1], Vincenzo Giannini[1], Peter Török[1], Tyler Roschuk[1] & Stefan A. Maier[1]

The absorption process of an emitter close to a plasmonic antenna is enhanced due to strong local electromagnetic (EM) fields. The emission, if resonant with the plasmonic system, re-radiates to the far-field by coupling with the antenna via plasmonic states, whose presence increases the local density of states. Far-field collection of the emission of single molecules close to plasmonic antennas, therefore, provides mixed information of both the local EM field strength and the local density of states. Moreover, super-resolution localizations from these emission-coupled events do not report the real position of the molecules. Here we propose using a fluorescent molecule with a large Stokes shift in order to spectrally decouple the emission from the plasmonic system, leaving the absorption strongly resonant with the antenna's enhanced EM fields. We demonstrate that this technique provides an effective way of mapping the EM field or the local density of states with nanometre spatial resolution.

[1] The Blackett Laboratory, Department of Physics, Imperial College London, London SW7 2AZ, UK. Correspondence and requests for materials should be addressed to E.C. (email: e.cortes@imperial.ac.uk) or to S.A.M. (email: s.maier@imperial.ac.uk).

Super-resolution microscopy techniques based on the localization of single molecule fluorescence have found extensive use in recent years in the fields of chemistry and biology, allowing for local probing and mapping of cellular structures. Towards this end, the primary techniques used are Stochastic Optical Reconstruction Microscopy (STORM) and Photo Activated Localization Microscopy (PALM), among others[1–5]. In these super-resolution approaches, the antigen-antibody strategy is commonly employed; thus fluorescent markers can report on the location of particular structures or regions of interest. Wide-field, laser illuminated, fluorescent microscopy is then performed on the sample. By forcing the majority of the molecules to be in temporary dark states[6], what would have been bright images composed of thousands of overlapping fluorescent point spread functions (PSF) that make up conventional fluorescence microscopy images, become sparse images where the individual PSF, corresponding to single molecules, are spatially separated. In this way, each PSF can be localized, returning the location of the free-space emitting molecule that generated the PSF to nm accuracy[7]. This process is cycled over many images of different active emitters. Once all of these positions have been localized, the resultant points can be plotted to produce a super-resolution image of the emitter distribution. These techniques have been hugely successful, starting a wave of new observations of biological and chemical structures and mechanisms[8–11]. Adapting these techniques, Cang et al.[12] recently proposed to extend the concepts of super-resolution fluorescence microscopy to study plasmon-induced electromagnetic (EM) fields by mapping the near-field inside 15 nm plasmonic hotspots on a rough metal surface.

Materials that exhibit plasmonic resonances allow for intense light focusing, thereby enabling EM energy transfer from the far to the near-field or vice versa[13]. When properly designed, plasmonic nanostructures can, therefore, act as optical nanoantennas and are key for the fabrication of devices capable of converting conventional photonic-scale optical fields to nanometre-scale volumes (producing EM hotspots)[14]. As these hotspots typically have dimensions on the order of only 10 s of nanometers, it is not possible to resolve and study them using conventional, diffraction limited, optical methods. Sub-diffraction approaches to the study of plasmonic materials and devices have provided valuable nm-scale information. Scanning near-field optical microscopy (SNOM) is one of the few optical techniques with sub-diffraction capabilities, with the resolution of an SNOM system being limited by the radius of curvature or aperture size of the probe, which can also cause non-trivial perturbations to the system[15]. Some other strategies aimed at nanostructure characterization have involved either using a single molecule fixed to the end of a probe as a constant source of illumination[16] or placing an optical antenna at a probe tip to map the directionality of the antenna's emission when scanning fluorescent molecules[17]. While these techniques have improved our ability to study fluorescence in plasmonic systems, they do not offer the ability to probe the far-field generated EM hotspots that are produced near plasmonic nanoantennas. Electron microscopy-based techniques, such as cathodoluminescence and electron energy loss spectroscopy, can be used to probe EM fields on these length scales; however, these techniques do not provide details of hot-spot-emitter interactions[18–21], and have in general severe support substrate constraints. Gaining knowledge of the complex light−matter interaction processes that occur when an emitter is placed in a sub-diffraction EM hotspot remains an active challenge in nanophotonics. Advances in this area would have uses in diverse fields such as (bio)sensing, non-linear optics, imaging and energy conversion, among others[22–25].

By using fluorescent molecules as near-field probes, Cang et al.[12] aimed to produce a direct map of the EM field without any interpretive complication. Plasmonic standing waves can give large near-field enhancement of EM fields but also lead to extreme fluorescence enhancement, especially within their hotspots[26]. By using the same localization technique that had previously been applied in STORM and PALM, and relying upon the fact that the intensity of the emission of a fluorescent molecule is proportional to the local near-field strength it experiences, it was thought to have been possible to map the EM fields with nm resolution. However, the fluorescent emission of these molecules is modified not just by their presence in the high intensity EM fields[26–28] but also by the effect of the plasmon resonance on the local density of states (LDOS)[29–33]. These enhancement effects that happen via the absorption process and the emission process of the dyes, respectively, can be expressed concisely as:

$$S = \frac{\eta}{\eta_0} \frac{|\boldsymbol{\mu_1} \cdot \mathbf{E}|^2}{|\boldsymbol{\mu_2} \cdot \mathbf{E_0}|^2} = \frac{\eta}{\eta_0} \frac{|\mu|^2}{|\mu|^2} \frac{|E|^2}{|E_0|^2} \frac{\cos^2(\theta_1)}{\cos^2(\theta_2)} \quad (1)$$

where $S$ is the total fluorescence enhancement for an emitter − antenna interaction, $\eta$ is the quantum efficiency (QE) of an interacting emitter, $\eta_0$ is the free-space QE, $\mathbf{E}$ and $\mathbf{E_0}$ are the enhanced and free-space electric fields at the illumination frequency, respectively, $\boldsymbol{\mu}$ is the dipole moment of the emitter and $\theta$ is the angle between the dipole moment and the electric field. Note that equation (1) expresses the general case of the total generated fluorescence light and does not include details of the collection efficiency of the system, which may be modified in the presence of an antenna, as shown later in the text. For optimally field aligned dipole moments we arrive at:

$$S = \frac{\eta}{\eta_0} \frac{|E|^2}{|E_0|^2} \quad (2)$$

By increasing the local EM field around an emitter, it will spend less time in its ground state before excitation, thus increasing the number of photons emitted in a given time period[34]. The effect of the increased LDOS on the emission often takes the form of a change in the QE of the molecule, that is, the molecule is able to emit into the optical states of the plasmonic antenna. If tuned correctly this can lead to an increase in QE and a decrease in the lifetime of the fluorescence, which can increase the amount of light emitted by the molecule over a given time period if quenching via non-radiative channels is avoided[35]. For high QE dyes, there may be no change in their QE when interacting with a plasmonic antenna; however, the emission is re-radiated via the antenna. Complications arise for localization methods when the light is re-radiated via plasmonic states[31,32,36–38]. For molecules that experience such coupling with the plasmonic structure, the localized position will typically not correspond to the real location of the molecule[17,37,39]. Rather, localized positions in these cases will be pulled towards the 'photonic centre of mass' of the system and away from the molecule's true location. Because most plasmonic resonances are spectrally broad in comparison to the absorption and emission bands of a standard fluorescent molecule, both processes are typically resonant with the antenna, making it difficult to accurately localize an emitter's position. The emission of a molecule next to a plasmonic system is then a complex process that cannot be simplified by the free-space emitting approximation as proposed in ref. 12, making the localization of these type of events not as straightforward as it is for the bio-systems using STORM, PALM and so on[37,40–42]. As recently reported by Darby et al.[43], the absorption processes at the single (few) molecule level in plasmonic systems can be modified by the presence of an antenna with important implications for molecular plasmonics, enhanced spectroscopies and so on. Similarly, the emission process of a single molecule in a plasmonic hotspot needs a greater understanding to fully address the relevant aspects of the re-emission of light to the far-field[44].

Here we propose to decouple the excitation and emission processes of an emitter in a hotspot through the use of molecules with large Stokes shifts. By employing pre-designed antennas—with fixed and well-known far-field induced hotspots—combined with polarization-sensitive super-resolution fluorescence microscopy, we demonstrate the differences on the localization of single-molecule events when diminishing the effects of emission-coupling to the plasmonic modes of the system. We achieve this by keeping the absorption process on resonance with the plasmon resonance of the antenna and by selecting from two types of dye molecules whose emissions are either on or partially off resonance with the available plasmonic states of the system. In this way we move back towards a quasi-free-space emission setting where the localization position again refers more closely to the real position of the probe, thus allowing effective EM-field mapping with nm resolution. Furthermore, we support these findings with Finite Difference Time Domain (FDTD) simulations. Finally, we compute and compare the localization and enhancement values for both dyes, disentangling the contribution of the EM field and the available LDOS on the absorption and emission processes for single emitters in a plasmonic hotspot. Through the sub-diffraction localization of emitters within an EM hotspot, important information can be extracted on the actual profile of the hotspot and its interaction with single emitters, enabling one to unravel this complex scenario of interactions, with implications in photonics and plasmonics[25].

## Results

**Near-field super-resolution localization microscopy.** Let us begin by describing the experimental approach to achieving single-molecule interaction with the plasmonic antennas that allows for the super-resolution localization of the molecule's emission events. Consider Fig. 1a. By putting a low concentration of our fluorescent probes (~5 nM) into dimethyl sulfoxide (DMSO) and introducing this onto an Al tri-disk antenna sample, Brownian motion moves the molecules about and allows them to interact with the sample surface at random locations, as shown schematically in Fig. 1a–i. By tuning the concentration and image capture integration time we are able to ensure that only a single molecule interacts with our antenna at any one time. While adsorbed at the surface, this molecule is effectively stationary and its emission produces a PSF in the far-field. The brightness of this PSF is proportional to the molecule's fluorescence, and an increase in the intensity of the PSF reflects an enhancement of molecular emission. The way that light is emitted from this hybrid molecule-plasmonic system, however, changes if the emission is on resonance (Fig. 1a–ii) or off resonance (Fig. 1a–iii) with the antenna. Figure 1b further illustrates this concept through FDTD simulations of the emission from a dipole placed on the right side of the antenna system. The emission of the dipole is tuned through different wavelengths while the resonance of the Al tri-disk antenna remains constant with a peak at ~400 nm. As the emission is tuned from completely on resonance to off resonance with respect to the LDOS peak of the system (blue line in Fig. 1b), less of the emission is coupled into the far-field via the plasmonic antenna. As mentioned, light that couples into the plasmonic antenna will re-radiate to the far-field differently than from the molecule alone (or from the off-resonant molecule situation)[37,40]. This causes a shift in the localized position of the molecule towards the system's photonic centre of mass. Finally, once desorbed the molecule leaves the area open for the next single molecule to adsorb (Fig. 1a–iv)[12]. The molecules are transient on the surface of the sample, and for the low laser powers employed in this experiments the total number of collected photons in each single molecule event is limited by the absorption desorption time of the fluorescent molecules more than by their photobleaching.

Over a long sequence of images, the interactions of thousands of emitters with the plasmonic antenna are observed and the localization process (Fig. 1c) is performed in each case. Full details of the localization method are documented in ref. 45. Briefly, for an emitting single molecule we first obtain a far-field image of the molecule's emission. From this image, a raw fluorescence PSF is obtained. In Fig. 1c–i,ii, the colour indicates the electron count at each camera pixel. A Gaussian is then fitted to this data, as shown in Fig. 1c–iii as a colour plot and c-iv as a wire mesh plot. This fit is optimized using a Maximum Likelihood Estimation method and the centroid position of the Gaussian is determined. This centroid position corresponds to the actual location of the emitting molecule in the free-space case. The overall precision of the localization method is primarily dependent on the number of photons in the PSF and the fluorescence background in the wide-field image. The solid colour profile plot in Fig. 1c–iv is a Gaussian whose full width half maximum (FWHM) is the precision of the Maximum Likelihood Estimation fit, which is shown below as a colour plot in Fig. 1c–v. Figure 1c, as a whole, demonstrates the change in scale from raw data to the localized position (further details about the localization method are available in the Supplementary Note 1). The ability of this method to accurately determine the position of an emitter itself is only valid when the emission of the molecule is not affected by coupling with the plasmonic antenna—a fact we examine experimentally later in this work.

The tri-disk arrangement used in this work is shown in the SEM image in Fig. 2c. The structure consists of three aluminium disks, 70 nm in diameter, spaced by 30 nm gaps. Antennas were fabricated, spaced by 2.5 μm, in arrays to avoid antenna–antenna interaction but allowing for the probing of multiple structures in a single wide-field imaging run (Fig. 2f). Samples were fabricated on glass cover slips using electron beam lithography (full details are provided in the Methods section and Supplementary Note 2). Aluminium was chosen to give us a plasmonic resonance primarily in the blue end of the visible spectrum[46], thus leaving free spectral space in the red/near-infrared region where the emission from large Stokes shift emitters can be located, and therefore diminishing the emission interaction with the plasmonic structure. DMSO was chosen as a solvent for our fluorescent molecules due to their excellent solubility in it and the high stability of Al antennas in this medium. The resonance spectrum of the Al tri-disk structure in DMSO is shown as the black curve in Fig. 2a (details on dark-field microscopy can be found in the Supplementary Note 3 and Supplementary Fig. 1). The resonance is centred at ~400 nm and, as is typical for Al antennas on a glass substrate, is quite broad[47]. We have verified by pre and post dark-field spectroscopy that the antennas remain unaltered over the course of our measurements. This is a critical detail, in particular for Al antennas that can be quite reactive[48]. A benefit of using Al is its formation of a native oxide layer of ~3–5 nm thickness, which acts to stabilize the antenna's surface[49]. This oxide also allows emitters to approach near to the surface without their fluorescence being quenched by the energy transfer mechanism[30,50]. This particular tri-disk arrangement was chosen to allow us to study how polarization and spectral decoupling of the emission change the localization of a molecule's emission and, by extension, changes the super-resolution maps of the surface. The tri-disk structure has gap resonances in more than one polarization direction; as such we can expect to see a considerable difference between the super-resolution localization maps obtained under different polarizations if the emission is reporting the EM-field enhancement (which is polarization sensitive). Conversely, if the emission occurs through a plasmonic coupling, the polarization sensitivity of the results will be diminished as this coupling is independent of the incident polarization of the light (further details on this aspect are discussed later in the text).

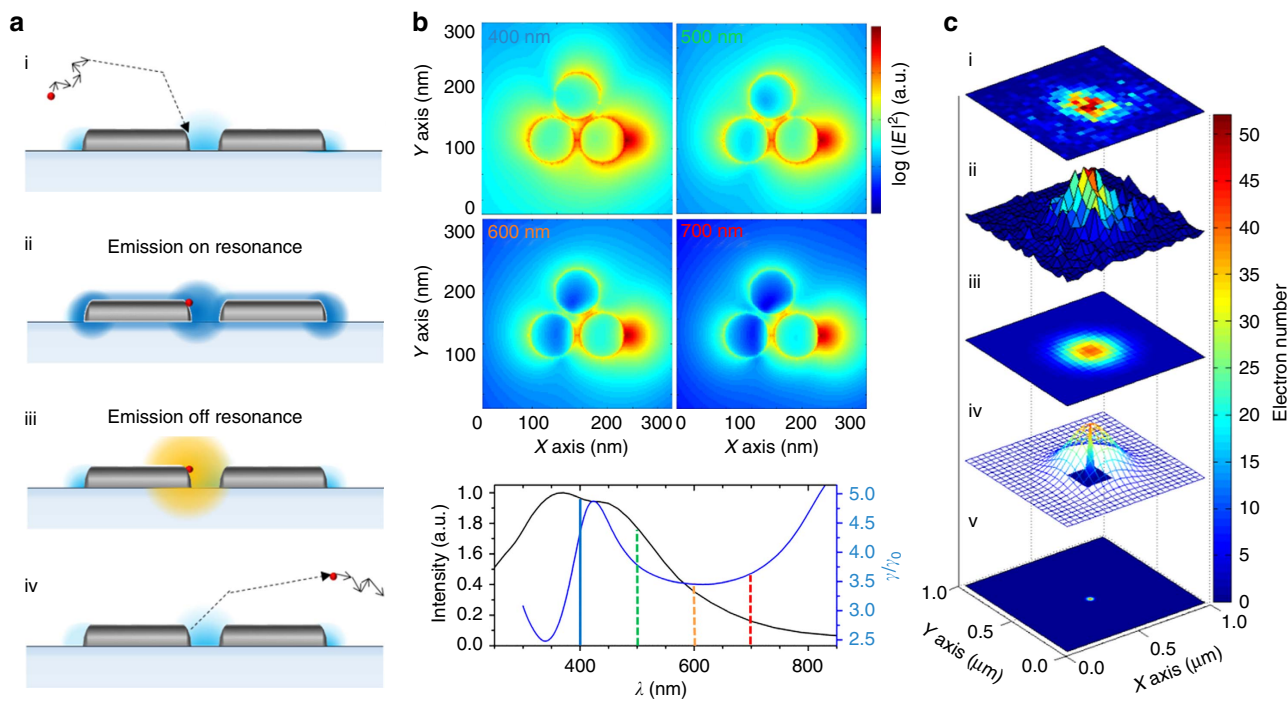

**Figure 1 | Interaction and localization of an emitter in a plasmonic hotspot.** (**a**) Scheme of the emitter interacting with the plasmonic hotspot. (i) A single molecular probe diffuses to the surface of an antenna via Brownian motion where it is adsorbed. (ii) For a double-resonant dye (absorption and emission on resonance with the antenna plasmon resonance and $\gamma/\gamma_0$ radiative enhancement contribution, respectively), light is emitted into the far-field directly from the molecule, and indirectly via the antenna (by coupling to the available modes in the plasmonic system)—leading to a delocalized position of the emission. (iii) Emission from a molecule for which only the absorption is resonant with the plasmonic mode. (iv) Once the probe is bleached and/or desorbs from the surface, it leaves the system free for a new probe molecule to arrive. (**b**) FDTD simulations of a dipole placed 10 nm to the side of a plasmonic Al tri-disk antenna emitting at different wavelengths. Emission from the dipole is tuned from an on-resonance condition (400 nm) to progressively more off resonance with respect to the LDOS peak of the system (blue line shown in the spectrum in the central bottom panel). The black line in the bottom panel is the scattering spectra of the Al tri-disk structure. (**c**) Super-resolution localization process for an emitting single molecule. (i) The EMCCD camera image (raw data) is taken and (ii) a surface plot of the raw data is produced and (iii) fit with a Gaussian contour. The centroid position (solid contour) of the Gaussian (mesh) contour is determined. The FWHM is the precision of the localization. (v) Finally, the localized position of the emission origin is recovered—for an uncoupled probe, this corresponds to the position of the molecule.

To study this, we use two very similar dyes, Pacific Blue (PB) and Pacific Orange (PO). The maximum absorption peak wavelengths of both PB and PO are both centred near 410 nm, shown in Fig. 2a as the blue and orange dashed lines, respectively. Both lines can be seen to be strongly on-resonance with the Al tri-disks. The emission peak values in DMSO for these dyes are at ∼450 and ∼575 nm, respectively. The emission of PO, therefore, shows a large Stokes shift, as shown in Fig. 2b, compared to PB. Overlapping radiative emission enhancement provided by our tri-disks with the emission peaks of our dyes, the black curve in Fig. 2b, shows that the emission of PB is further within a strong emission enhancement band of the system. As such, one expects that the emission from PB will couple more strongly to the plasmon modes of the tri-disk system. This idea is shown conceptually in Fig. 2d, which illustrates an example of the emission from a resonantly coupled dye and Fig. 2e, which shows emission from a decoupled one. In the coupled case, the emitter takes on the emission profile of the simple metal antenna[17].

The sample was mounted onto an inverted microscope system with a total internal reflection (TIR) illuminator. A schematic of the optical setup is shown in the Supplementary Fig. 2. The sample was coated with the dye solution and illuminated with a 405 nm laser in a TIR configuration, which helps to reduce the background fluorescence due to diffusing molecules by only illuminating the structures via an evanescent field at the sample surface. We use lower powers to maintain the molecules in a linear response regime, even when factoring in the enhanced local fields that arise due to our antennas (see Methods section for further details). The laser illumination is filtered out and the sample fluorescence is imaged using an EMCCD camera with an exposure time of 100 ms. During the imaging sequence, scattered laser light from the antenna is monitored via a CCD camera. Using the scattered light from a fixed reference point on the sample, the sample position and focus are corrected in real time via a piezoelectric stage. This allows for the collection of extended image sequences over several hours without any defocus or drift, ensuring the accuracy of our super-resolution maps (details on the focus-lock implementation are provided in the Supplementary Note 4).

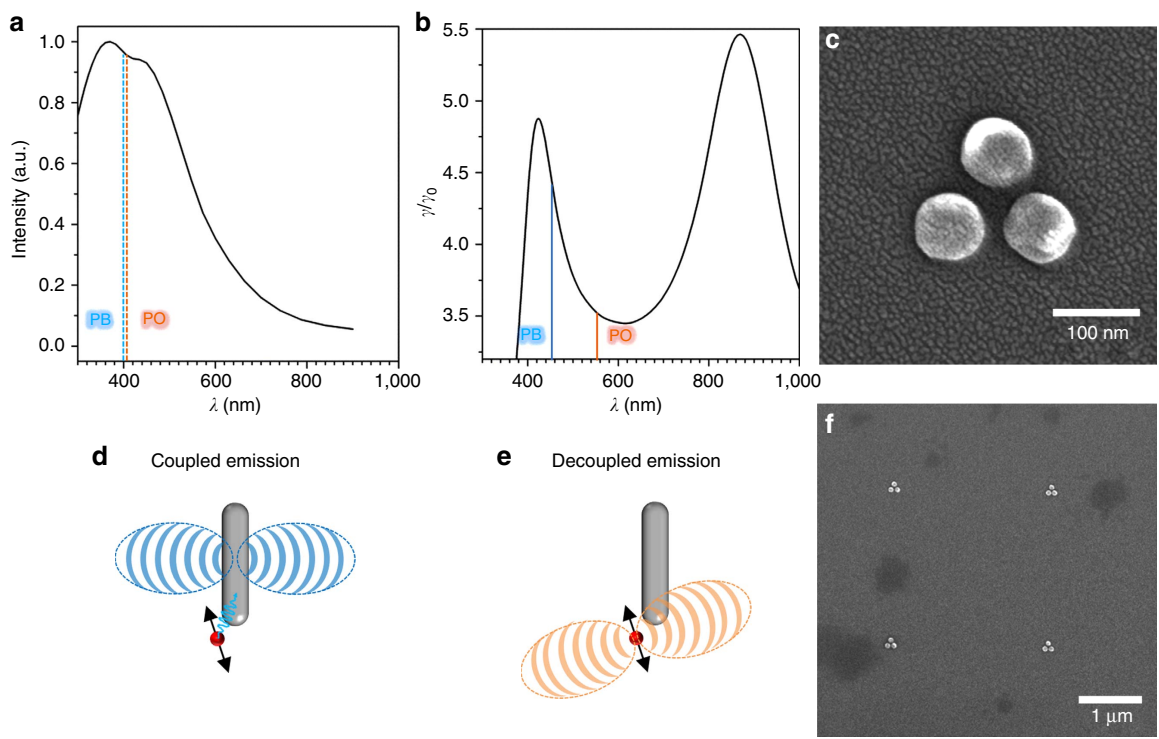

**Figure 2 | Spectral characteristics of the dyes used in this work and a schematic of their expected behaviour next to a plasmonic antenna.** (**a**) Simulated scattering spectra of an Al tri-disk system in DMSO (black) overlaid with the maximum absorption wavelengths for PB and PO dyes (blue and orange dashed lines, respectively). (**b**) $\gamma/\gamma_0$ radiative enhancement of the system overlaid with the maximum emission wavelengths for PB and PO dyes (blue and orange lines, respectively). (**c**) SEM image of the Al tri-disk antenna. (**d,e**) Scheme showing the expected difference for coupled (that is, PB) and decoupled (that is, PO) emission from an equivalently positioned dipole in proximity with a simple plasmonic antenna. (**f**) SEM image showing four antennas from the Al tri-disk antenna-array.

**Polarization-resolved localization maps**. Figure 3a shows the super resolution localization map obtained for the double-resonant, absorption and emission, PB dye coupled to our tri-disk structure and illuminated with a 405-nm laser. The polarization of the illumination source is indicated by the red arrow and each 'pixel' in the image corresponds to an area of 10 nm × 10 nm (more details about the localization field maps are provided in the Supplementary Note 5 and Supplementary Fig. 3). Figure 3c shows FDTD simulations of the near-field electric-field distribution for both polarizations. It should be noted here that the slight asymmetry in the y-polarized near-field electric distribution is due to the TIR illumination geometry. In the case of x-polarized light, one can see that an electric field enhancement is produced between the bottom two disks. When these results are compared to the x-polarization results in Fig. 3a, one readily sees that this gap enhancement is reflected in the intensity increase in this gap region for the tri-disk nanoantenna. When the illumination is switched to y-polarized light, the total fluorescence intensity for PB decreases but the localization map remains very similar in its overall spatial distribution compared to the x polarized one. The antenna-emitter coupling is independent of illumination polarization as it depends on the orientation of the molecule, which we have no control over. However, the EM field strongly depends on the polarization of the source. Obtaining very similar spatial distributions in the localization maps for both polarizations confirm that the free-space approximation is not valid for an emission-coupled dye: the invariance of the localization maps with respect to the polarization is caused by the re-radiation of the emission to the far-field via the antenna. If the localization maps were reporting the EM field distribution, we would expect a strong spatial dependence on polarization.

Figure 3d shows a simulation of the effect of the available LDOS on the emission enhancement, $\gamma/\gamma_0$, for a tri-disk nanoantenna for $\lambda = 450$, 575 and 700 nm. This simulation was done via calculation of the dyadic Green's function, **G**, for the electric field at a position **r** due to an x-polarized point source at $\mathbf{r}_0$:

$$\bar{\bar{\mathbf{G}}}_x(\mathbf{r}, \mathbf{r}_0) = \frac{\mathbf{E}(\mathbf{r}) c^2 \varepsilon_0 \varepsilon_r}{\omega^2 \mu} \quad (3)$$

for the x component of the Green's function $\bar{\bar{\mathbf{G}}}_x$, where $\varepsilon_0$ is the permittivity of free-space, $\varepsilon_r$ is the dielectric constant of the medium, $\mu$ is the dipole moment of the emitter and $\omega$ is the angular frequency of the dipole. The Green's function takes the form of a 3 × 3 matrix whose elements correspond to dipoles oriented along the Cartesian directions. From the Green's function we can calculate a nanoantenna's effect on the density of states available to the dipole, which in turn allows us to calculate the emission enhancement (further details are provided in the Fluorescent Emission Enhancement Simulation section in the Supplementary Note 6).

At 450 nm, the emission wavelength of PB, a large portion of the emitted light gets radiated to the far-field via the nanoantenna. As such, the localization for the coupled case of PB and our tri-discs cannot be relied upon to provide the true location of the molecule. Rather, the localizations reflected in our super-resolution maps are a complex combination of the light radiated from the molecule in a hotspot and light radiated via the antenna. Disentangling these two effects is not easily accomplished. Attempts to return the true molecular position, via simulation, for a simple plasmonic system have been recently proposed[37]. An examination of the emission enhancement plot at 575 nm indicates a greatly reduced effect on the antenna-coupling phenomena at that wavelength (Fig. 3d). As such, one expects that the localization of PO, which emits at this wavelength, will primarily reflect the EM hotspot's contribution to fluorescence enhancement. By simply decoupling the emission we should be able to minimize unnecessary complexity and interpretive issues with the results.

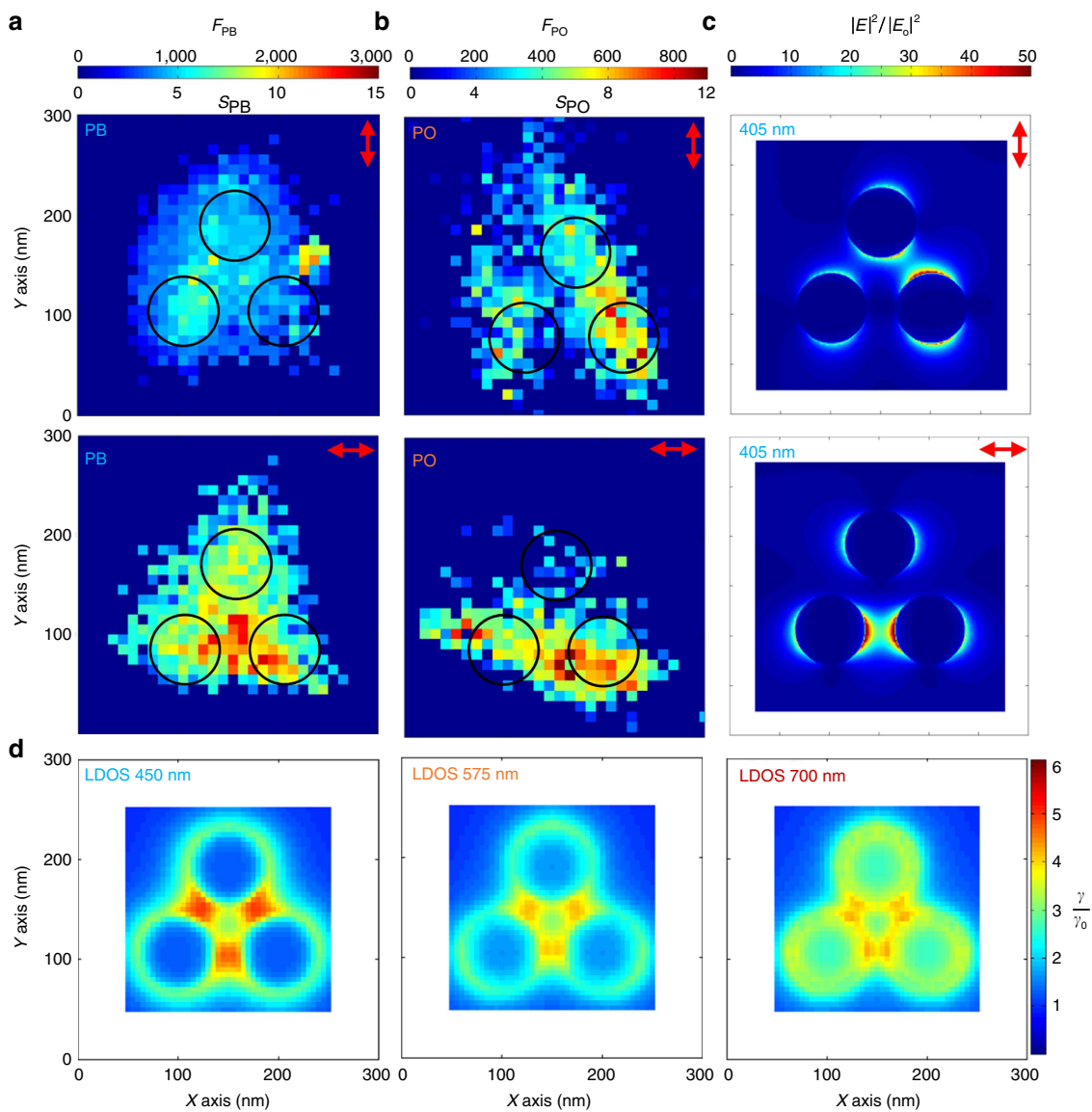

**Figure 3 | Super-resolution localization maps with EM field enhancement and emission enhancement results for a tri-disk system.** Experimental results for the localization fields for Al tri-disks structures illuminated by a 405-nm laser in TIR with the polarization orientation indicated by the red arrow. The colour corresponds to the fluorescent enhancement level $S$ and total collected fluorescence $F_T$ for a molecular probe at that apparent position for the cases of (**a**) Pacific Blue dye and (**b**) Pacific Orange. (**c**) FDTD simulation of the $|E|^2/|E_o|^2$ near-field distribution for TIR illuminated Al tri-disk structures on a glass substrate in water (**d**) FDTD simulations of the effect of the increased LDOS on the emission enhancement of a dipole at each pixel position at wavelengths of 450, 575 and 700 nm.

With these points in mind, Fig. 3b shows the localization maps for $x$ and $y$ polarized light obtained for PO on the tri-disk sample. Notice that the total collected fluorescence scale, $F_{TPO}$, is $\approx 1/3$ that of $F_{TPB}$; a detail we will revisit later in this paper. Also to be noted is the $x-y$ broadening of the PO localization map. This is due to the localization position beginning to converge towards the true molecule position and away from the 'photonic centre of mass' of the antenna system. This agrees with the results of Wertz et al.[37], as molecules located ~90 nm from the antenna are able to re-radiate via the plasmonic system when coupling with the plasmonic states; thus leading to a mislocalization of the emitters' position from the actual position of the antennas, as observed for PB but not for PO. In these PO maps, it is also readily apparent that the localization results show distinct features when the illuminating polarization is rotated—reflecting the polarization dependence of the EM field enhancement observed in the simulation results (Fig. 3c), with a strong correlation to the gap enhancements expected under these polarization conditions. This shows that as the molecule's fluorescent emission moves away from the plasmonic resonance, the localization field map becomes more sensitive to the polarization change of the illumination field. This reinforces the fact that once the number of available plasmonic states to couple into has been reduced, the localization maps reflect primarily the EM field distribution.

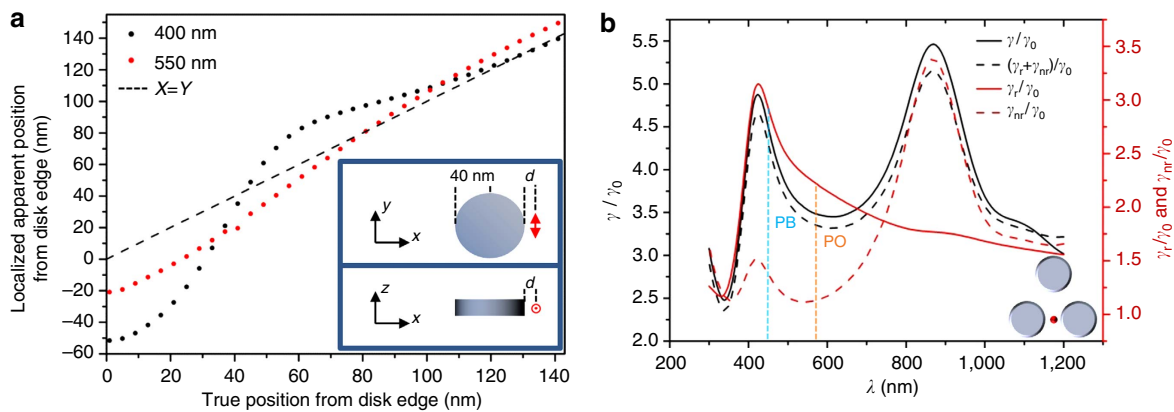

**Figure 4 | Super-resolution localization improvement and spectral response of emission enhancement factors.** (**a**) Apparent localized position with respect to the true position for a dipole emitting at 400 nm (black dots) compared to one emitting at 550 nm (red dots) for an Al disk. The dashed line corresponds to perfect localization. As the emission of the dipole is moved towards a lower $\gamma/\gamma_0$ radiative enhancement contribution, the error in the localization is notably reduced. (**b**) For a combined $x-y$ polarized dipole located at the position marked in red on the insert we show the total emission enhancement (black dashed line), radiative enhancement (red solid line) and non-radiative enhancement (red dashed line). Finally we simulate the total emission enhancement effects via calculation of the dyadic Green's function for the structure (black solid line). Note from the spatial distribution of the emission enhancement (Fig. 3d), placing the dipole in another hot-spot of the structure would give the same result.

**Fluorescence gain calculations.** After analysing the effects of the LDOS and the EM field on the localizations events for both dyes, let us now compute their respective contributions to the total fluorescence gain values for each dye. We will use this analysis to demonstrate that the localization of PO reasonably represents the EM field around our structures.

As we have mentioned, when the emission wavelength of the dye overlaps with the peak in the LDOS due to the plasmonic structure, leading to an enhanced radiative emission rate, there is a shift of the emission position towards the photonic centre-of-mass of the system that leads to a shift of the localization position. The extent of the mislocalization is illustrated in Fig. 4a, which demonstrates the apparent localized position for a molecule as a function of its position from the edge of a single Al disk. The dashed line shows the case of perfect localization. For a dye molecule emitting at 400 nm (black dotted line), where there is a strong coupling associated with the available plasmonic states—reflected in the increased LDOS at that wavelength—the mislocalization is strong. At 550 nm (red dotted line), however, the emission is partially decoupled from radiating via the plasmonic structure, as can be seen in Fig. 4b, and the apparent position of the molecule approaches its real position (Fig. 4a).

First, note that although plasmonic structures present a large increase in the number of states for photons to couple into, these states will have both radiative and non-radiative pathways for the out-coupling of this energy[26]. Figure 4b shows the enhancement factor for the emission due to an increase in the LDOS, as well as for the individual radiative and non-radiative contributions for an Al tri-disk antenna. As was seen in Fig. 3d, the radiative enhancement via plasmonic coupling for the Al tri-disk antenna is greatly reduced when moving from 450 to 575 nm. Over this wavelength range, these radiative contributions (solid red line in Fig. 4b) are the dominant factor to the overall emission enhancement, with only a small contribution from non-radiative pathways (dashed red line). Moreover, one can see that the emission enhancement increases at 700 nm, a result not readily expected based on the plasmonic resonance of the antenna alone (see Fig. 2a). This increase is, however, related to an increase of the non-radiative channels for $\lambda > 650$ nm, which correspond to inter-band transitions in Al (further details on the wavelength dependence of the LDOS are provided in the

Supplementary Figs 4–7). We note that although radiative effects are not completely eliminated at 575 nm (the PO emission wavelength), going above 650 nm in the emission of the dye would lead to significant coupling to non-radiative states. This coupling does not affect the localization process, as these states are dark to our observation efforts; however, the brightness of the interactions would be greatly quenched. As such, 575 nm provides a good compromise between these two situations. An ideal case—that is, complete decoupling—might never be possible to achieve for plasmonic systems; however, as we have shown, measurable differences in the localization maps can be observed by reducing/minimizing it.

Since we are using a predesigned antenna geometry producing a known hotspot we are able to compare the experimental values of fluorescent enhancement in Fig. 3a,b to the expected values in order to better understand the source of the emission enhancements. Furthermore, we take information from the maximally enhanced dyes at each spatial location (that is, when $\boldsymbol{\mu}$ is aligned with the electric field); hence $\boldsymbol{\mu}$ can be simplified in equation (1) and from equation (2) it can be seen that the total fluorescence enhancement $S$ is made up of two components: the enhancement of the QE, $f_\eta = \eta/\eta_0$, due to the antennas effect on the LDOS, and the E field enhancement, $f_E = |E|^2/|E_0|^2$. The expected new QE for each dye can be calculated via equation (see Supplementary Note 7 for the derivation of this equation):

$$\eta = \frac{\gamma_r/\gamma_0}{(1-\eta_0) + \gamma_r/\gamma_0 + \gamma_{nr}/\gamma_0} \qquad (4)$$

Values for the radiative enhancement $\gamma_r/\gamma_0$ and non-radiative enhancement $\gamma_{nr}/\gamma_0$ are taken from Fig. 4b. For the following analysis, we use $\gamma_{rPB}/\gamma_0 = 2.9$, $\gamma_{nrPB}/\gamma_0 = 1.35$, $\gamma_{rPO}/\gamma_0 = 2.2$ and $\gamma_{nrPO}/\gamma_0 = 1.2$. The intrinsic QE for PB is $\eta_{0PB} = 0.78$ and for PO can be estimated as $\eta_{0PO} \approx 0.5$, respectively[51,52]. From equation (4), we then arrive at new values for the QEs of $\eta_{PB} = 0.65$ and $\eta_{PO} = 0.56$ when interacting with the tri-disk antenna. This yields QE enhancement factors of $f_{\eta PB} = 0.83$ and $f_{\eta PO} = 1.13$. Both of these enhancements are near unity. $f_E$, however, is $\approx 20$ for the largest part of the field distribution for both dyes, as they are under the same illumination fields inside the antenna. We are, therefore, in a $f_E$-dominated regime as $f_E \gg f_\eta$. This is supported by the observed $S$ values in Fig. 3a,b ($S_{PB} \approx 16$, $S_{PO} \approx 13$), which are in line with the expected values. Because of this the fluorescent enhancement values in Fig. 3a,b are directly related to the electric field enhancement of the antenna and not combined with a large QE enhancement factor as would be the case for a low intrinsic QE dye[26,53].

We can therefore conclude that although plasmonic coupling produces misleading localization results, the resulting emission enhancement has nearly no effect on the total number of collected photons when employing high QE dyes[13,26]. For this reason, we get similar fluorescence gain values for both dyes, even for PO whose emission is not in resonance with the plasmonic antenna. Because of this we would expect the total fluorescence collected for each dye to be similar as in both cases the illumination is the same. However, as we noted earlier the light collected for PB is ~3X that of PO. To analyse the absolute fluorescent values and to determine the source of this discrepancy, let us begin by using equation (2) and defining the total enhanced fluorescence $F_T$ for a single molecule interacting with a plasmonic system as:

$$F_T = SF_0 \quad (5)$$

where $F_0$ is the intrinsic fluorescence for a free-space emitting dye molecule and can be written as:

$$F_0 = A_{eff}(\lambda)\sigma_{abs}(\lambda)\eta_0|\mu|^2|E|^2 \quad (6)$$

here $\sigma_{abs}$ is the absorption cross-section for a single molecule and $A_{eff}$ is the collection efficiency of the optical system. Similarly to equation (2) this assumes field-aligned molecules. Combining equations (2, 5 and 6) and taking the ratio of the total enhanced fluorescence for PB and PO dyes we can write:

$$\frac{F_{TPB}}{F_{TPO}} = \frac{\eta_{0PO}}{\eta_{PO}}\frac{\eta_{PB}}{\eta_{0PB}}\frac{|E_0|^2}{|E|^2}\frac{|E|^2}{|E_0|^2}\frac{A_{effPB}\sigma_{absPB}\eta_{0PB}|\mu_{PB}|^2|E|^2}{A_{effPO}\sigma_{absPO}\eta_{0PO}|\mu_{PO}|^2|E|^2} \quad (7)$$

As we are using the same plasmonic system for both dyes we can cancel the electric field terms, thus arriving at:

$$\frac{F_{TPB}}{F_{TPO}} = \frac{\eta_{PB}}{\eta_{PO}}\frac{A_{effPB}\sigma_{absPB}}{A_{effPO}\sigma_{absPO}} \quad (8)$$

Inserting the values for the QEs we calculated from equation (4) we obtain $\eta_{PB}/\eta_{PO} = 1.15$. For our optical setup the largest contribution to $A_{effPB}/A_{effPO}$ is the ratio between the QE of the EMCCD camera at 450 and 575 nm; 0.85 and 0.95, respectively. This yields $A_{effPB}/A_{effPO} \approx 0.89$. Combing these results, the difference between fluorescence intensity for each dye comes almost entirely from the change in $\sigma_{abs}$. The total number of collected photons is, from Fig. 3a,b, $F_{TPB} \approx 3,000 > F_{TPO} \approx 900$. The ratio of these values is in agreement with that obtained from the literature values of the absorption coefficients $\sigma_{absPB} \approx 46,000\,M^{-1}cm^{-1}$ and $\sigma_{absPO} \approx 25,000\,M^{-1}cm^{-1}$ (ref. 51). $\sigma_{abs}$ for single molecules is slightly modified when adsorbed on the antenna, which likely produces the small deviations we see in our results[43].

Through this analysis, we have shown that the increased LDOS presented by the plasmonic structure does not affect the gain values significantly due to the high initial QE of PB and PO. However, the coupling to the plasmonic modes of the system still prohibits the correct localization of PB (that is, it affects the spatial location of the emission, without affecting the QE of the emitter itself). Because the enhancement of the QE for PO is near unity, the gain values measured predominantly reflect the EM field enhancement, thus our super-resolution localization maps represent an accurate optical mapping of the EM field around a plasmonic antenna.

## Discussion

The complex interplay between the absorption and emission processes at the single molecule level in the presence of a plasmonic antenna makes mapping of enhanced fields difficult. By decoupling these processes using a dye with a large Stokes shift, we have demonstrated that we can minimize the plasmonic coupling, which is a major source of error in the localization of single-molecule events when performing super-resolution localization fluorescence microscopy in plasmonic systems. Activating different hotspots in the antenna by changing the polarization allowed us to show that the output data contained in the fluorescence enhancement maps is the combinations of the EM field (polarization sensitive) and the increased LDOS accessible by the molecule (which is independent of the polarization of the illuminating light). By employing an emitter with a large Stokes shift we minimized the number of states available to couple into. This enabled significantly more accurate mapping of the enhanced EM fields alone than has been possible to date. Moreover, our use of high QE fluorescent dyes produced a situation where the plasmonic coupling can affect their apparent emission position, without the increase LDOS producing a strong enhancement to the total emission. This allows us to easily link the fluorescence enhancement values with the EM field enhancement only. A secondary example illustrating the ability to resolve—and map the EM fields around—two distinct dual disk antennas within a sub-diffraction hotspot is provided in the Supplementary Note 8 and Supplementary Fig. 8.

Further to this current work, we expect that by analysing the difference between multiple dyes this method could give information on how the LDOS changes with different emission wavelengths. Using techniques similar to this we envisage the possibility for a direct LDOS probing method using active fluorescent molecules or quantum dots, with the peak of the LDOS designed to overlap only with the emission wavelength of these probes[31,54,55]. We expect this approach will provide a more accurate method of optically probing EM hotspots in plasmonic systems and will help to yield a better understanding of the fundamental processes taking place when an emitter interacts with a plasmonic nanostructure. Finally, by providing a method for the reliable localization of single molecules when interacting with a plasmonic antenna, we also expect to increase interactions between the fields of nanoscopy and plasmonics. Nanoscale information can be accessed for plasmonic systems by employing this approach.

## Methods

**Sample fabrication.** Samples were fabricated on No.1 glass cover slips (VWR). Before fabrication, the cover slips were rinsed with acetone, IPA and DI water. A thin film (200 nm) of poly(methyl methacrylate) (PMMA), was formed onto the glass surface via spin coating at 3,500 r.p.m. for 60 s, followed by a 5-min baking step at 160 °C. A conductive layer of ESPACER 300Z was then spin coated onto the sample (1,500 r.p.m. for 60 s followed by a 60 s, 100 °C bake). Nanostructures were then patterned into the PMMA via electron beam lithography (Raith e-line). After pattering, the espacer was removed via submersion in de-ionized water. The pattern was then developed in an IPA:MIBK mix (3:1) for 1 min followed by a cleaning plasma ash etch step to improve metal adhesion (Electronic Diener Femto, 7 s at 40% power). The samples were then coated with 30 nm of Al via thermal evaporation (Angstrom A-mod). The sample was completed with lift-off step in acetone.

**Numerical analysis.** To better understand molecular interaction with our plasmonic system various finite-difference time-domain simulations were conducted using the finite-element Maxwellian equation solver Lumerical. Simulations were split into three groups: (a) EM near-field study for TIR illumination simulation of a glass/water interface surface of tri-disk antenna using 50° plane wave illumination at 405 nm. (b) Scattering cross-section spectra of tri-disk antenna using direct plane wave illumination on a glass sub-straight in DMSO. (c) LDOS study using Green's function analysis of a dipole emitter tri-disk antenna interaction.

**Dark field microscopy.** Samples were side-illuminated with a Nikon Intensilight C-HGFI mercury lamp. High angle scattered light is then collected by an M-Plan APO NUV 50X, NA 0.42 objective, which is transmitted via a FG600AEA Thorlabs fibre to a Princeton Instruments spectrometer.

**Super-resolution mapping.** The samples were illuminated using a 405-nm laser diode source (Coherent Cube) using an inverted microscope fitted with a TIRF illuminator (Nikon) and a ×100 oil immersion objective (NA 1.49, Nikon). The laser light was filtered using dichroic (Z405rdc Chroma) and emission filters (ET420LP Chroma). Single molecule fluorescence was collected using an EMCCD camera (Photometric Evolve 512). Each frame had a 100 ms exposure time, with ~30 ms of dead time between acquisitions. SR maps were constructed from a minimum of 60,000 images and, on average, contain ~4,000 successful localization points after suitable filtering. In order to account for sample drift scattered laser light from the sample was reflected by the dichroic mirror and collected via a second camera (QICam). A disk in each array of structures was used as a reference point and the scattered laser light was localized and used to correct the sample position.

During the measurement process, the illuminating laser was kept at low power (on the order of ~$10^{-1}$ W cm$^{-2}$) to ensure that in the presence of an enhanced EM field around our plasmonic structures, our dyes continued to operate in a linear response regime. The reader should note that this is one to several orders of magnitude less than conventional super-resolution microscopy techniques and as a result unenhanced molecules at the glass/sample interface are not observed.

**Fluorescent dyes.** Pacific Blue succinimidyl ester (PB) and Pacific Orange succinimidyl ester triethylammonium salt (PO) were purchased from Thermo Fisher Scientific and were used as received. Dilutions in DMSO were characterized by extinction spectroscopy and concentrations of 5–10 nM were used. Attempts to use dilutions in water-based buffers led to a complete degradation of the Al antennas. In the same way we observed that neutral charged molecules and anions (PB and PO, respectively) allow strong single molecule interaction by using the Brownian motion approach. Similar attempts with cationic dyes led to very weak interaction rates.

**Localization code.** Raw data are filtered to discard image frames that do not contain active molecules. These filtered data are then localized using Maximum Likelihood Estimation method with the aid of code distributed with the work by Mortensen et al.[45] Localized data are then filtered for spurious points, due to failed localization, by intensity, location and variance in an effort to remove localization due to multiple fluorescent molecule interaction events and PSF distorting effects as documented in Su et al.[56] Data were then sectioned into 10 nm bins. The average of the highest intensity results are taken from each bin and plotted.

**Data availability.** All relevant data are available from authors on request.

### Acknowledgements
D.L.M. acknowledges the support of the Leverhulme Trust. E.C. is supported via a Marie-Curie Horizon 2020 Fellowship. The authors acknowledge funding provided by grants from the Engineering and Physical Sciences Research Council Reactive Plasmonics Programme (EP/M013812/1), the Leverhulme Trust, the Royal Society and the Lee-Lucas Chair.

### Author contributions
T.R. and S.A.M. conceived the experiment and P.T. refined necessary details of the optical setup and performance. E.C. and D.L.M. devised the method of decoupling the excitation and emission enhancements and determined suitable dyes to perform the experiments. D.L.M., P.T. and T.R. designed and constructed the microscopy system as well as developing the real-time tracking setup. D.L.M. implemented the code to perform the real time tracking. D.L.M., T.R. and E.C. fabricated the samples. D.L.M. and E.C. performed the experiments. D.L.M. and V.G. designed the simulations and D.L.M. performed them. T.R., P.T. and S.A.M. supervised all aspects of the work; E.C. supervised the experiments. E.C. and D.L.M. drafted the manuscript and all authors contributed to its revision.

### Additional information
**Supplementary Information** accompanies this paper at http://www.nature.com/naturecommunications

**Competing financial interests:** The authors declare no competing financial interests.

**Reprints and permission** information is available online at http://npg.nature.com/reprintsandpermissions/

**How to cite this article:** Mack, D. L. *et al.* Decoupling absorption and emission processes in super-resolution localization of emitters in a plasmonic hotspot. *Nat. Commun.* **8,** 14513 doi: 10.1038/ncomms14513 (2017).

**Publisher's note**: Springer Nature remains neutral with regard to jurisdictional claims in published maps and institutional affiliations.